\documentclass[aps,preprint]{revtex4-1}

\usepackage{amssymb,amsfonts,amsmath}
\usepackage{graphicx}
\usepackage{xcolor}              
\def\lett#1{(\textbf{#1})}

\begin{document}

\title{Disease-induced resource constraints can trigger explosive epidemics}

 \author{L. B\"{o}ttcher}
 \affiliation{ 
 ETH Zurich, Wolfgang-Pauli-Strasse 27, CH-8093 Zurich,
Switzerland}

\author{O. Woolley-Meza}
 \email{owoolley@ethz.ch}
 \affiliation{Computational Social Science, ETH Zurich, Clausiusstrasse 37, CH-8092 Zurich,
Switzerland}

 \author{N. A. M. Ara\'ujo}
 \affiliation{Departamento de F\'{\i}sica, Faculdade de Ci\^{e}ncias, Universidade de Lisboa, P-1749-016 Lisboa, Portugal, and Centro de F\'isica Te\'orica e Computacional, Universidade de Lisboa, Avenida Professor Gama Pinto 2, P-1649-003 Lisboa, Portugal}
   
 \author{H. J. Herrmann}
 \affiliation{
 ETH Zurich, Wolfgang-Pauli-Strasse 27, CH-8093 Zurich,
Switzerland, and Departamento de F\'isica, Universidade
Federal do Cear\'a, 60451-970 Fortaleza, Cear\'a, Brazil}

 \author{D. Helbing}
 \affiliation{Computational Social Science, ETH Zurich, Clausiusstrasse 50, CH-8092 Zurich,
Switzerland}

%
%
%
%

\begin{abstract}

Advances in mathematical epidemiology have led to a better understanding of the risks posed by epidemic spreading and informed strategies to contain disease spread. However, a challenge that has been overlooked is that, as a disease becomes more prevalent, it can limit the availability of the capital needed to effectively treat those who have fallen ill. Here we use a simple mathematical model to gain insight into the dynamics of an epidemic when the recovery of sick individuals depends on the availability of healing resources that are generated by the healthy population. We find that epidemics spiral out of control into ``explosive'' spread if the cost of recovery is above a critical cost. This can occur even when the disease would die out without the resource constraint. The onset of explosive epidemics is very sudden, exhibiting a discontinuous transition under very general assumptions. We find analytical expressions for the critical cost and the size of the explosive jump in infection levels in terms of the parameters that characterize the spreading process. Our model and results apply beyond epidemics to contagion dynamics that self-induce constraints on recovery, thereby amplifying the spreading process.
\end{abstract}
\maketitle

\section*{Introduction}
The global mobility of individuals that sustains modern economic activity has increased the risk of pandemics~\cite{Hufnagel:2004, Vespignani24072009}. For example, approximately 3.1 billion passengers travel by air every year, accelerating disease spread and making spreading patterns more complex~\cite{Colizza06,Brockmann13}.
The biggest pandemic on record, the 1918 Spanish influenza, was responsible for the death of 50 million people at a time when global mobility was slower and much less.
Thus, while the containment and mitigation of global disease spread in our time has become harder due to high interconnectedness, it could save millions of lives, making it one of the most pressing 21st-century challenges.
Important progress in understanding epidemic spreading has been made using mathematical models that capture the underlying processes and their dependence on
infection and recovery rates~\cite{Anderson91,Dorogovtsev08,Barrat08}.
These studies have focused mainly on the interplay
between the dynamics and the structure of
interactions~\cite{Kuperman01,Pastor-Satorras01,Newman02,Belik11,Dickison12,Brockmann13}, and on identifying the main spreaders~\cite{Colizza06,Kitsak10,Balcan10}. Progress has also been made in 
developing strategies to mitigate epidemic
risks, for example through vaccination~\cite{Pastor-Satorras01,Pastor-Satorras02,Cohen01,Ferguson06,Epstein09,Schneider11,vandenBroeck11}.
Many studies address the inherent limitations in the resources available to treat or prevent diseases, for instance by finding the optimal way to allocate a limited pool of immunization doses and antivirals ~\cite{Colizza07}. This work assumes that the resources available to treat and prevent disease are independent of the evolution of the epidemic. However, the health burden a disease places on a population can limit the production of these resources~\cite{Helbing13}. Here we investigate a simple model that includes such disease-induced resource constraints.

A healthy and productive population is needed to enable the recovery of those that become ill: Healthy individuals are both the human resources needed to provide health services and pharmaceuticals, and contributors to the healthcare budget through taxes and insurance premiums.
These mechanisms operate on shorter and longer timescales but the availability of resources is always most relevant for diseases where the cost of treatment is high. This can apply to diseases that are endemic in a population and have depressed the economic productivity and growth of a region over a sustained period, leading to an underfunded health care system that cannot provide the necessary level of treatment. For example, it has been estimated that countries where Malaria was widespread in 1965 had a 1.3\% lower annual growth rate than other countries between 1965 and 1990~\cite{gallup2001economic,sachs2002economic}. Similarly, AIDS could lead to the collapse of many African economies within a few years~\cite{bell2003long}.
On the other hand, resource constraints also apply to sudden outbreaks that evolve too fast for the hospital and drug production systems to adapt the available resources to supply sufficient treatment in the necessary time frame. A timely example are highly virulent emergent pathogens, such as SARS and H1N1 influenza, which are widely feared because of their potential to generate the next great pandemic~\cite{mclean2005sars,webby2003we}. An even more recent example is the Ebola epidemic unfolding in West Africa, where one of the main initial challenges was the high infection rate in the already limited population of nurses and doctors, due partly to understaffing and suboptimal protective practices at treatment centers \cite{pandey2014strategies,merler2015spatiotemporal}.

To investigate the effect of disease-induced resource constraints on epidemic spreading we consider the simplest epidemiological model that captures longer time scales, the \textit{Susceptible-Infected-Susceptible} (SIS) model. This model applies directly to infections that do not confer immunity, such as many bacterial infections, but also to the general case where a population is exposed to a number of diseases, and individuals that have been infected by one disease may easily become infected by another. It is also a good approximation for cases where immunity is conferred but it wanes rapidly relative to the infection dynamics.
The specific model we use applies to cases where no external resources can enter the system  -- either the disease is so widespread that there are no unaffected regions from which resources can be transferred, or transfer cannot occur for logistical or political reasons.
We emphasize that our goal is not to capture the detailed dynamics of specific diseases, but rather to understand what type of behavior can emerge due to disease-induced resource constraints, and to characterize analytically how this behavior depends on the basic parameters of a contagion model. We also seek to determine the generality of our findings, and will discuss some cases where simulations or analytic arguments indicate that qualitatively similar behavior arises independent of model assumptions.

\section*{Model and Methods}


In  the SIS formalism individuals are either in a susceptible state or in an infected state. We will denote by $s(t)$ and $i(t)$ the fraction of susceptible and infected
individuals, respectively. Susceptible individuals are healthy but
become infected at a characteristic infection rate $p$ when interacting
with infected individuals. Infected individuals recover
and become susceptible again at a recovery rate $q$. In this model individuals recover independently of available resources. 

\begin{figure*}
\includegraphics[width=0.5\columnwidth]{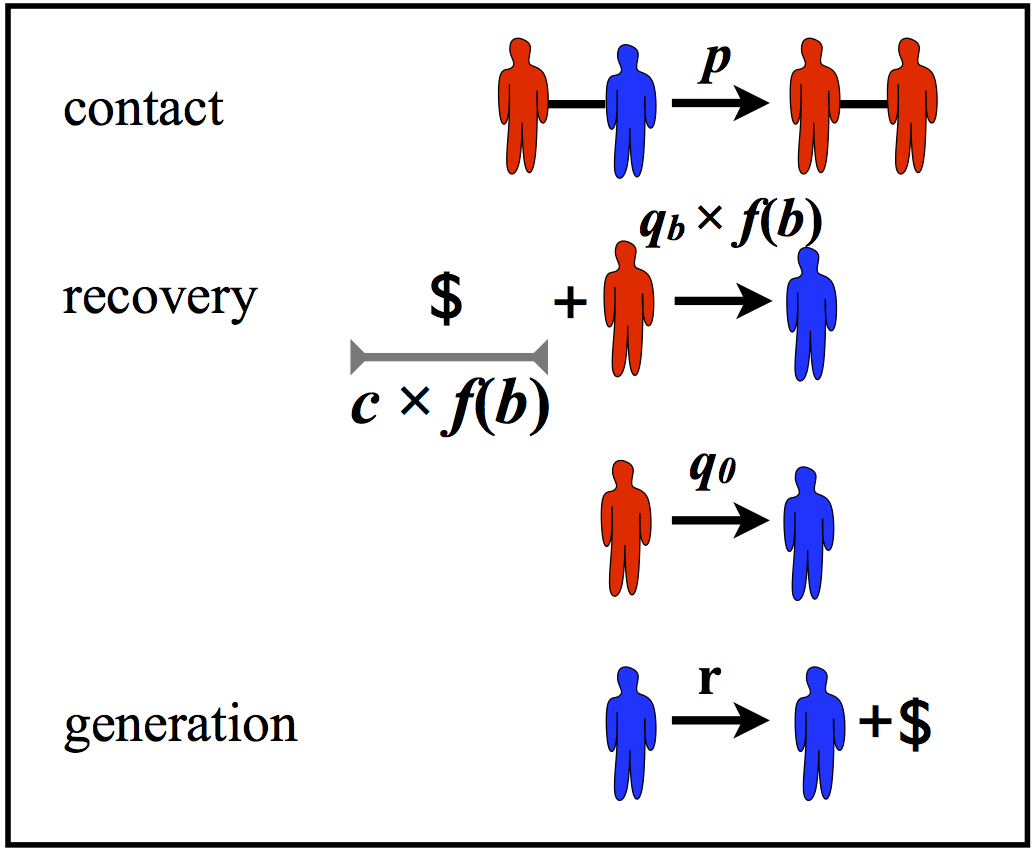}
\caption{\textbf{The bSIS model.} The four transitions that describe the bSIS model, each occurring at the rate denoted above the transition arrow. As in the SIS model, infected (red) individuals contact and infect susceptible (blue) neighbors at rate $p$. Infected individuals will recover at a resource mediated recovery rate $q_bf(b)$ if the necessary resources $cf(b)$ are applied, and with an additional baseline rate $q_0$, independent of resource availability. A unit of budget (denoted by the \$ sign) is generated by each susceptible at a rate $r$. Without loss of generality we have normalized the model so that one unit of budget is generated by every individual per unit of time and thus $r=1$. (Little men courtesy of Alejandro Morales Gallardo.) \label{fig::bSIS}}
\end{figure*}

To model a recovery process that is dependent on healing resources generated by the healthy population, we introduce our budget-constrained Susceptible-Infected-Susceptible (bSIS) model illustrated in Fig.~\ref{fig::bSIS}. This model modifies the SIS dynamic by (1) introducing a \emph{global} budget $b$ that increases by one unit with the number of healthy individuals per unit time and (2) constraining recovery, at least partially,
if the necessary budget to cover costs of healing is not available. In other words, there is a background recovery process that is independent of the budget available, and this process occurs at a rate $q_0$. However, to reach the full potential for recovery, resources must be invested. Specifically, a cost $c$ is incurred \emph{for each individual} that recovers through treatment. Thus, this model captures the most general case, defining costs relative to the resource unit generated per unit time. Although we refer to every individual producing a unit of budget per unit time, this is identical, formally, to the situation where only a fraction of the population contributes to the budget, as long as the contributing population is just as likely as the rest of the population to become infected and to recover. Clearly the rate of resource mediated recovery will depend on the total budget $b$ available. 
If there is no budget, there will be no resource mediated recovery. On the other hand, with a positive budget some recovery is possible, and the rate of recovery increases  until the maximum resource mediated recovery rate $q_b$ is reached. To express this dependence we introduce the budget function $f(b)$ which satisfies $f(b)\equiv 0$
for $b\leq0$ and $0<f(b)\leq1$ for $b>0$. Thus, the resource mediated recovery rate, given budget $b$, is just $q_bf(b)$ as shown in Fig.~\ref{fig::bSIS}.
	
The simplest budget function fulfilling the required conditions is the Heaviside step
function $\Theta(b)$, which is one for $b>0$ and zero otherwise. In fact, for the sake 
of analytical tractability, this is the function we will consider for most of our analysis. However, we will show in the \textit{Generalized Behavior} section that the key results we derive using this specific budget function hold under any budget function that satisfies the general conditions defined above.

\subsection*{Mean-field approximation}
To define the spreading dynamics we need to specify the interaction process between infected and susceptible individuals. One common starting point, which yields powerful insight and is analytically tractable, is the mean-field approximation. Topologically, the mean-field limit corresponds to an uncorrelated random graph, 
where every individual interacts with $k$ randomly selected others (i.e. neighbors), to which it can transmit the disease. With this assumption the dynamics of the fraction of infected individuals are given by
\begin{equation}\label{eq::infected}
\frac{d i(t)}{d t}=kpi(t)s(t)-[q_0 + q_bf(b)]i(t)\, ,
\end{equation}
where
$s(t)=1-i(t)$, so that the rate equation for the fraction of susceptibles is
analogous.
The first contribution on the right-hand side
accounts for the infection of susceptible individuals that are in contact with
infected ones, assuming an infection rate of $p$. The last term accounts for both the background and the budget dependent recovery processes.

To fully describe the evolution of the mean-field bSIS system we couple Eq.~(\ref{eq::infected}) with the equation for the rate of change in the budget $b(t)$:
\begin{equation}\label{eq::budget}
\frac{d b(t)}{d t}=s(t)-c q_b f(b) i(t) \, .
\end{equation}
Here the first term accounts for the budget produced by healthy individuals, and the second term for the reduction in budget due to successful application of healing resources.
Note that setting $f(b)\equiv 1$ for all $b$ we recover the pure SIS system with recovery rate $q=q_0+q_b$ (in this case Eq.~(\ref{eq::budget}) is redundant). 

One of the main insights obtained from the SIS model without budget constraints, is the existence of a transition from a \textit{healthy} regime where the fraction of infected individuals vanishes in the thermodynamic limit (infinite system size), to an \textit{epidemic} one characterized by a non-negligible fraction of infected individuals~\cite{Barrat08}. 
The details of the transition depend on the interaction structure between individuals~\cite{Parshani10,Boguna13}. In the mean-field system the basic reproduction number $\tau=kp/q$, which is just the number of secondary infections generated by one infected individual in a population of susceptible individuals, fully characterizes the onset of the SIS epidemic transition. Specifically, the system is in the epidemic regime when $\tau$ is greater than the critical value
$\tau^*=1$~\cite{Marro99}, and the transition
to the epidemic regime is continuous. Furthermore, the level of infection in the epidemic regime increases with the  basic reproduction number. Importantly, the steady state infection levels in the population are independent of the initial condition in such a model (as long as the initial infection level is greater than zero).
We shall see in what follows, the introduction of a budget-limited recovery process transforms these dynamics by modifying the steady state infection level and introducing a dependence on initial conditions.


\subsection*{Numerical simulations on a social network}

In real systems disease transmission occurs through non-random interactions, in contrast to the assumptions of the mean-field approximation.
The interactions can be modeled as networks generated through the interplay between multiple
social mechanisms~\cite{Watts98,Palla05}. 
These generating mechanisms together lead to complex structural features that are not generically found in random networks, most importantly: high clustering
coefficients~\cite{Easley10}, community structure \cite{newman2011communities,Fortunato10}, and the small-world property~\cite{Watts98}.
Because the dynamics of disease spread can be highly sensitive to the interaction structure, we study the limited resource case on a real social network alongside the mean-field approximation. Specifically, we consider a school friendship
network shown in Fig.~\ref{fig::school_main}a, which exhibits the features characteristic of real social networks: clustering, community structure and the small-world property. This network was constructed based on the answers of school
students in the United States to an Add-Health
questionnaire~\cite{Gonzalez07}.  From the full set of networks obtained
from this data set, we consider the largest connected component, which consists
of $2539$ nodes and $20910$ edges. (See \textit{Supplementary Information} for a more detailed characterization of the network.) We will show that, surprisingly, the simple mean-field approximation  captures the behavior on this more realistic interaction structures between individuals.

\section*{Results}
\subsection*{The emergence of explosive epidemics }

We investigate the behavior of the
bSIS model on the static friendship network, considering a parameter regime where an epidemic arises in the SIS model ($\tau>\tau^*$). (All numerical results on the friendship network
are averages over $3200$ stochastic simulations.) To build intuition we initially consider the simplest case, where \emph{all} healing requires a budget contribution ($q_0=0$), and in addition we assume a Heaviside step function for the budget. Thus, unless otherwise specified, $q=q_b$ in the results that follow.
However, in the \textit{Generalized behavior} section we will show that the key characteristics in the behavior also hold qualitatively when there is background healing and for more general functions.

\begin{figure*}
\includegraphics[width=\columnwidth]{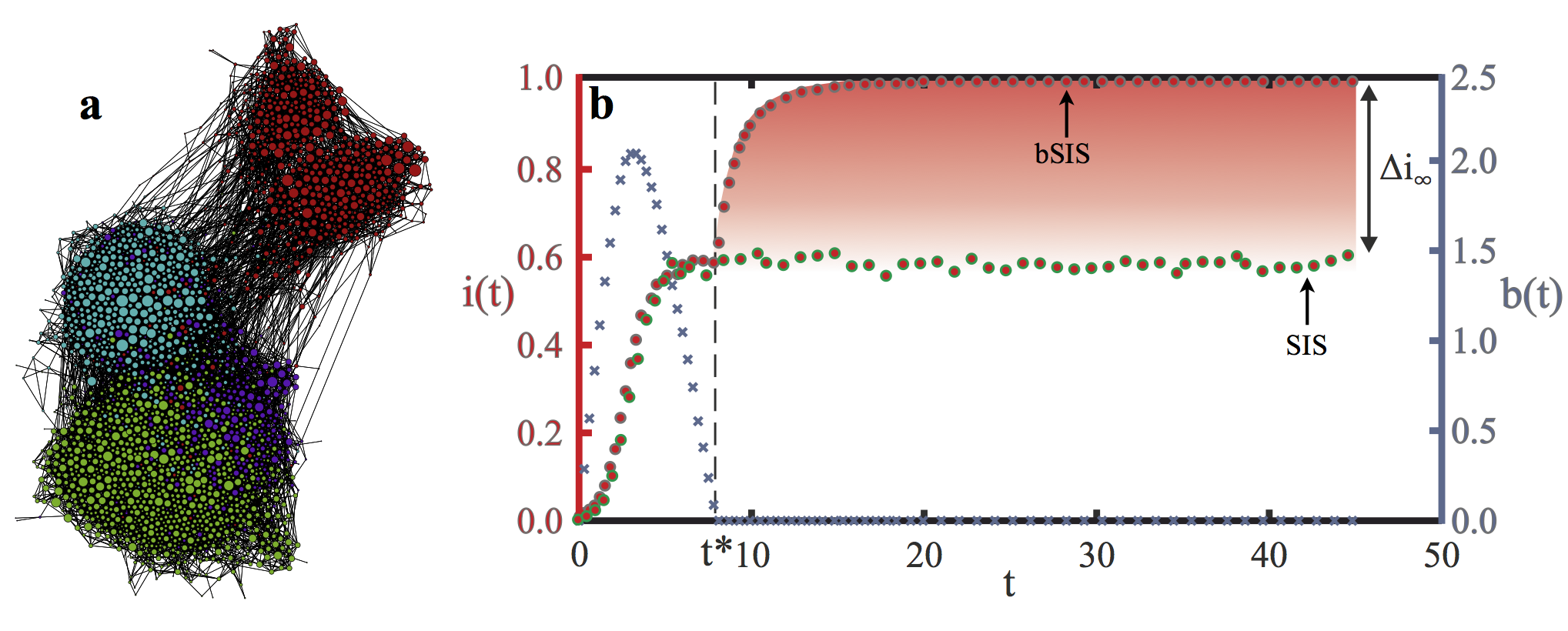}
\caption{\textbf{Epidemics self-accelerate when resources are endogenously constrained} \lett{a} The school friendship network, which exhibits long-range connections, clustering and community structure (colors denote communities found through modularity maximization). \lett{b} Evolution of
the fraction of infected individuals $i(t)$ (red circles) and the 
budget $b(t)$ (blue crosses), for healing costs above the
critical cost ($c=2>c^*\approx0.833$), on the friendship network. Recovery requires a budget: $q_0=0$ and $q_b=0.8$, and $p=0.285$. In the pure SIS model the fraction of infected individuals reaches a
stationary regime. The bSIS model also initially converges to the same
stationary regime until the budget is exhausted at a critical time
$t=t^*\approx 7.3$. For $t>t^*$ recovery is not possible anymore,
and the infection spreads to all of the population (i.e. $i(\infty)=1$). The difference in the steady state infection in both models, which is also the size of the jump in the bSIS model due to the budget constraints, is denoted $\Delta i_{\infty}$. 
\label{fig::school_main}}
\end{figure*}

In the early stages of the epidemic simulation we observe exactly the SIS behavior, as seen in Fig.~\ref{fig::school_main}b. We will refer to this, budget unrestricted infection incidence level, as the \emph{low} epidemic.
However, after a critical time  $t=t^*$ the epidemic spreading rapidly escalates---a situation that we describe as an ``explosive epidemic''. We will refer to this increased incidence level as the \emph{high} epidemic.
We see in Fig.~\ref{fig::school_main}b, an explosive epidemic where the entire
population is infected in the steady state ($i(\infty)=1$).

\begin{figure*}
\includegraphics[width=\columnwidth]{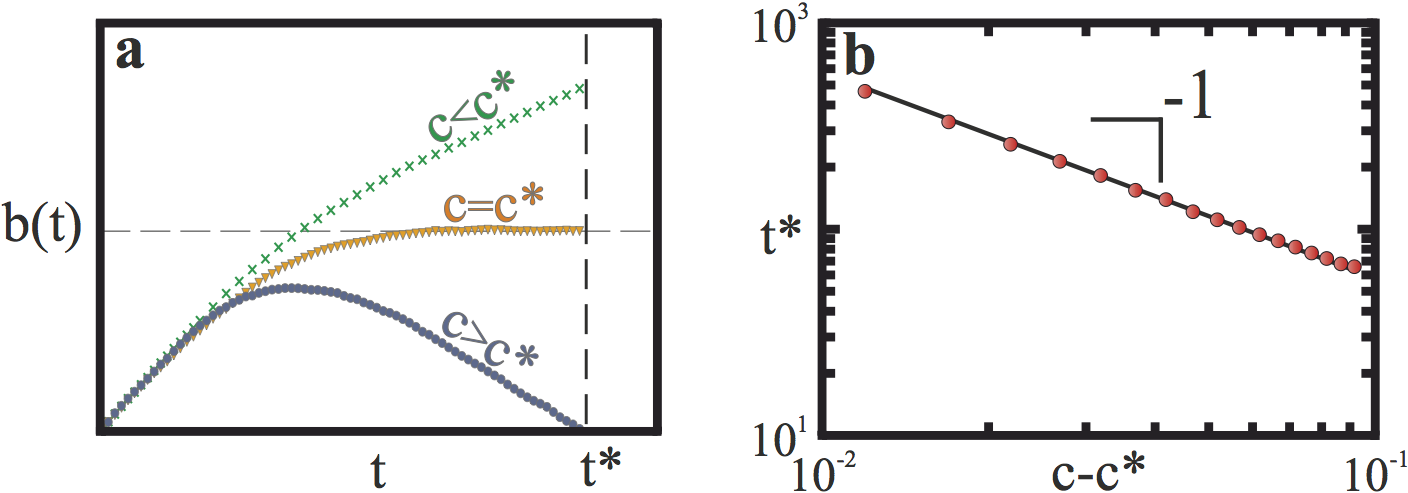}
\caption{\textbf{The time dynamics of the budget according to the cost of recovery.} 
\lett{a} Three different regimes of time-dependent budget development $b(t)$, depending
on the healing cost $c$. (Other parameters are fixed at values $q_0=0$, $q_b=0.8$, $p=0.285$.) For healing costs $c$ below the critical cost, $c^*$
(green crosses $c=0$), the budget is a monotonically increasing function
of time. Above the critical cost ($c=2>c^*$, blue circles), the budget
reaches a maximum and then decreases until it is exhausted at a critical time
$t=t^*$, allowing the infection to spread explosively. At the critical cost ($c=c^*= 0.833$, orange inverted triangles), the budget
saturates at a non-zero value. \lett{b} The
critical time scales with the distance to the critical cost
according to $t^*\sim(c-c^*)^{-1}$ (averaged over
$3600$ samples). Parameters are fixed at values $q_0=0$, $q_b=0.8$, $p=0.287$.
\label{fig::school_time}}
\end{figure*}

What leads to this drastic alteration of the SIS steady state infection level when we consider disease-induced budget constraints? The explosive epidemics arise when the healing budget is depleted and the normal recovery rate cannot be sustained. Depending on the healing cost $c$, one of the three
qualitatively different behaviors of the budget in time (see Fig.~\ref{fig::school_time}), will be instantiated.
When healing costs are small, the
budget systematically increases and the system evolves towards
an infinite budget. On the other hand, with large enough healing costs, the budget is used
up at a finite time $t=t^*$. After this time, any contribution to the budget by healthy
individuals is immediately spent on the recovery of only a small fraction of all the infected
individuals. Thus the epidemic spread outcompetes recovery, further decreasing the number of healthy individuals who can contribute to the budget. Eventually this process leads to a fully infected population. 
In between these two regimes, there is a critical cost ($c=c^*$) at which the budget saturates at a non-zero value. Here the infection level evolves as it would without a budget constraint, until it eventually exhausts the budget. Thus, integrating Eq.~(\ref{eq::infected})  we recover the evolution of infection time:
\begin{equation}\label{eq::infected_solved}
i(t)=\frac{i(0)\left(1-\tau\right)}{\left(1-\tau\left[1-i(0)\right]\right)e^{(1-\tau)t}-i(0)\tau}
\ \ ,
\end{equation}
where $i(0)$ is the initial fraction of infected individuals and we have assumed $f(b)=\Theta(b)$.

Similarly, to approximate all three budget behaviors as observed in Fig.~\ref{fig::school_time}a we can simply integrate Eq.~(\ref{eq::budget}) without a budget constraint ($f(b)\equiv1$):
\begin{equation}\label{eq::budget_solved}
b(t)=\frac{q\left[1+c\left(q-kp\right)\right]t+\left(1+cq\right)\log{\left[i(t)/i(0)\right]}}{kp}\ ,
\end{equation}
where $i(t)$ is given by  Eq.~(\ref{eq::infected_solved}).
(Here we have set the initial condition $b(0)=0$, without loss of generality.)

The crucial point is that we can find a critical cost $c^*$ above which positive contributions by healthy
individuals to the healing budget are insufficient to meet the costs generated by the
recovering patients (i.e. $s(t)<cqi(t)$). Thus,  according to Eq.~(\ref{eq::budget}), the budget will start decreasing until it is depleted in finite time and we are in the second case described above, where infection takes over the population.
Clearly, for a fixed cost and recovery rate, the resource demands for recovery are larger when there are more infected individuals. This means that the critical cost is determined by the maximum infection level that the system will reach before the budget starts to decrease, $\max_t[i_{\not \text{b}}(t)]$, where $i_{\not \text{b}}(t)$ denotes the infection level without a budget restriction.
Putting all of this together, in the mean-field case we can find $c^*$ by solving for $c$ at the stationary state of the budget, that is we set $d b(t)/d t=0$ in Eq.~(\ref{eq::budget}) with the constraint that the steady-state infection level must be  $\max_t[i_{\not \text{b}}(t)]$.
This gives
\begin{equation}\label{eq::critical_cost}
c^*=[kp \max_t[i_{\not \text{b}}(t)]]^{-1} \, .
\end{equation}

In the case we consider in Fig.~\ref{fig::school_main} the basic reproduction number is above the critical level (i.e. $\tau>1$), and $0<i(0)<1-\tau^{-1}$, which means that the pure SIS dynamic reaches the maximum infection level at the epidemic equilibrium $i(\infty)=1-\tau^{-1}$.
This gives the special case of Eq.~(\ref{eq::critical_cost}) 
\begin{equation}
 c^*=\left(kp-q\right)^{-1} \,. 
 \end{equation}
This case applies when $\tau>1$, as long as the initial infection level is not larger than the SIS epidemic equilibrium (i.e. $0<i(0)\leq1-\tau^{-1}$).
If the initial level of infection $i(0)$ is larger than the pure SIS equilibrium level then $\max_t[i_{\not \text{b}}(t)]=i(0)$, whatever the value of $\tau$. We can see this by noting that infection cannot increase above $i(0)$ if $b>0$, since the pure SIS dynamics apply and in the SIS model infection monotonically converges to the equilibrium. Thus Eq.~(\ref{eq::critical_cost}) implies that in this case the critical cost depends on the initial condition. In fact, at the critical cost $c=c^*$ the recovery demands must immediately match the generated resources. Since the initial budget is $b(0)$,
the infection level will oscillate around the initial infection level $i(0)$ as the budget continuously switches on and off.
This indicates that it is critical to treat diseases before the fraction of infected people escalates above the epidemic level since otherwise the epidemic will become explosive more easily (i.e. $c^*$ will be smaller). 

Interestingly, this last case implies that even for
$\tau\leq1$, where no epidemic phase is predicted in a pure SIS model, in the bSIS model we find a finite
critical cost above which infections spread
without efficient healing (dependent on the initial condition $i(0)>0$).
Furthermore,
the outcome of a disease outbreak in the SIS model can be fully characterized by the basic reproduction number $\tau=kp/q$, but in the
bSIS model two parameters are needed: $kp$ (the effective contact rate) and $q$ (the recovery rate). These represent
the time scales of infectious contact and recovery, respectively.  While the
dynamics of infection are still described by $\tau$, the budget
evolution depends on the competition between the time scales of recovery
and budget generation, thereby requiring one additional parameter.

Can we determine the critical time at which the budget is exhausted, given the cost of treatment? We would expect that the larger the healing
costs, the faster the budget is exhausted. We do in fact observe in our simulations that the critical time $t^*$ scales with
the healing cost as $t^*\sim(c-c^*)^{-1}$ (see Fig.~~\ref{fig::school_time}b).
Using Eq.~(\ref{eq::budget_solved}) we can determine $t^*$ by solving for $b(t^*)=0$ ( given $c>c^*$). 
When $\tau\leq1$ by definition $i(t^*)=i(0)$ and therefore Eq.~(\ref{eq::budget_solved}) implies $t^*=0$.
For the case $\tau>1$, if we assume $t^*\gg 1$, transients decay and the equilibrium epidemic infection level $i(\infty)=1-\tau^{-1}$ is reached without budget constraints. In this case
\begin{equation}\label{eq::critical.time}
t^*= \log{\left(\frac{i(\infty)}{i(0)}\right)}\tau
c^{*2}\left[(c-c^*)^{-1}+\frac{1}{\tau c^*}\right] \, .
\end{equation}
Thus,  $t^*\sim(c-c^*)^{-1}$ as found in the friendship network. If
$\tau\leq1$, one obtains $t^*=0$ as a consequence of the initial
condition $b(0)=0$, which implies that the treatment is constrained by
the available budget from the beginning. Whatever the contributions of healthy
individuals to the budget, they cannot cover the treatment costs.

\subsection*{Explosive epidemic transitions are discontinuous}

\begin{figure*}

\includegraphics[width=\columnwidth]{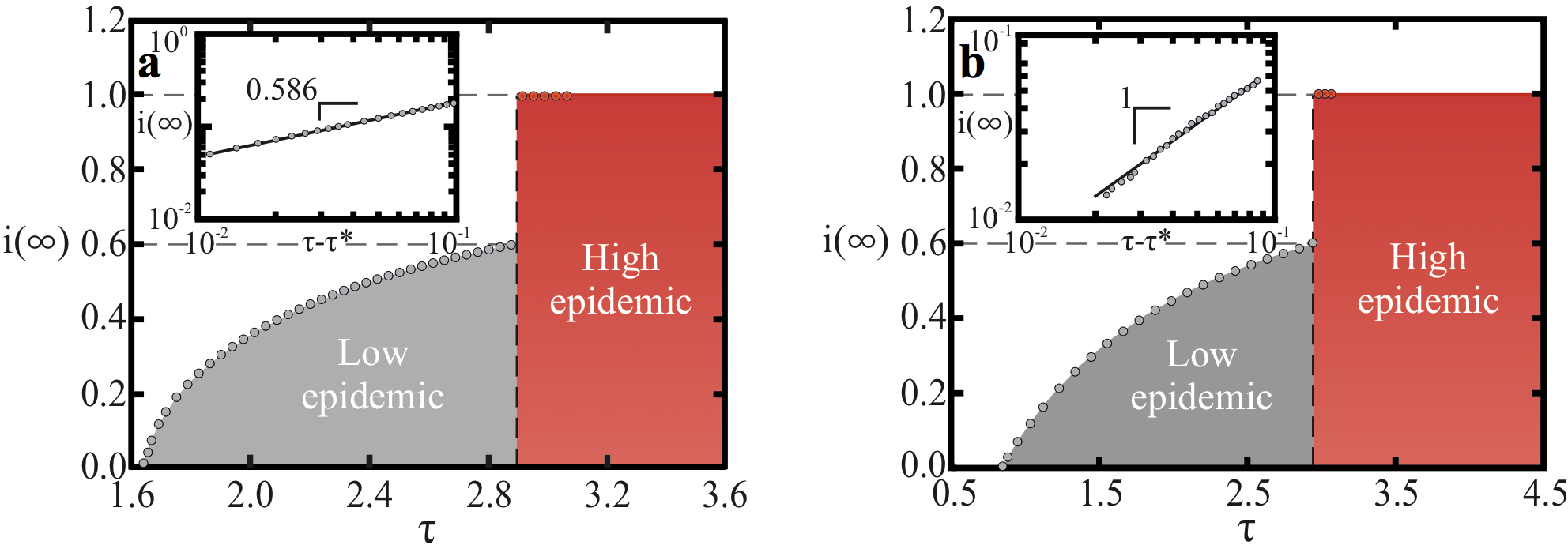}
\caption{\textbf{Discontinuous transition from the low to the high epidemic
regime.} \lett{a} Asymptotic fraction of infected individuals $i(\infty)$ on a
square lattice with $1024^2$ sites (we consider a von Neumann neighborhood, which has $k=4$ neighbors), for fixed costs $c=0.833$, and
recovery rate $q_b=0.8$, as a function of $\tau$ (we vary $p$, and hold $k$ and $q_b$ fixed). A discontinuous transition from the low (grey) to the high epidemic
regime (red) is observed. The inset shows the approach of
$i(\infty)$ to $\tau^*=1.6488\pm0.0001$, confirming the predicted
power-law scaling. The black solid line is a guide to the eye,
corresponding to $(c-c^*)^\beta$, where $\beta=0.586\pm 0.014$ is the
critical exponent of the order parameter for the SIS model on the square lattice.
The values of $\tau^*$ and $\beta$ where obtained from
Ref.~\cite{Moreira96}. \lett{b} Same as in (a) but on the
friendship network, where the average number of neighbors is $<k>=8.2355$. The inset shows that in the epidemic regime $i(\infty)\sim(\tau-\tau^*)$, just as in the mean-field limit of the SIS model~\cite{Marro99}. (Results are averages over $3600$ samples.) \label{fig::itaufn}}
\end{figure*}

The dynamic change at $t^*$, and the consequent drastic increase in the
fraction of infected individuals, is a continuous process in
time (see Fig.~\ref{fig::school_main}b).
However, the transition between the regimes is discontinuous in the model parameters
(the healing costs as well as the recovery and infection rates). For example, as illustrated in Fig.~\ref{fig::itaufn}, the transition is always discontinuous in the basic reproduction number $\tau$. The transition is also discontinuous in the sense that there is a jump, $\Delta i_{\infty}$, in the infection level (see Fig.~\ref{fig::school_time}b). We discuss below how the jump size scales with the model parameters.  The discontinuity of this jump in the final infected fraction $i(\infty)$ has important implications for the resilience of the health system and the control of disease, since it implies that a minute change in the properties of the disease or average contacts of an individual can abruptly shift the infection levels from a low equilibrium to one that is higher and harder to control. 

\subsection*{Explosive epidemics arise across network structures}

It is surprising that we can capture most of the dynamics of the epidemic on a real social network using only mean-field calculations. In particular, the same
critical exponents are found for the scaling of both the fraction of infected individuals 
and the critical time with the distance from the critical cost.
This leads us to ask whether the results we have found above are independent of network structure.
We hypothesize that there is a resemblance between the dynamics in the random network of the mean-field approximation and the real friendship network, because the long-range interactions present in both networks dominate the dynamics. To test this hypothesis we study
the bSIS model on a square lattice, where there are no long-range connections, and we would therefore expect to find different scaling
exponents if our hypothesis is true.
As shown in Fig.~\ref{fig::itaufn}a
we again find that the transition between the low and high epidemic regimes
is discontinuous. However, in the epidemic regime
$i(\infty)\sim(\tau-\tau^*)^\beta$ with $\beta\approx0.59$. This value
is instead consistent with the order parameter exponent $\beta=0.586\pm0.014$
reported for the classical SIS on the square lattice~\cite{Moreira96}.
However, we find that the critical time scales with the distance from the critical cost according to $t^*\sim(c-c^*)^b$, with $b=-0.98\pm0.03$,
which is numerically consistent with the one found in the mean-field
limit (see \textit{Supplementary Information}). This suggests that the effects of the cost of recovery on the system dynamics are similar across network topologies.

\subsection*{Generalized behavior}

\begin{figure*}
\includegraphics[width=\columnwidth]{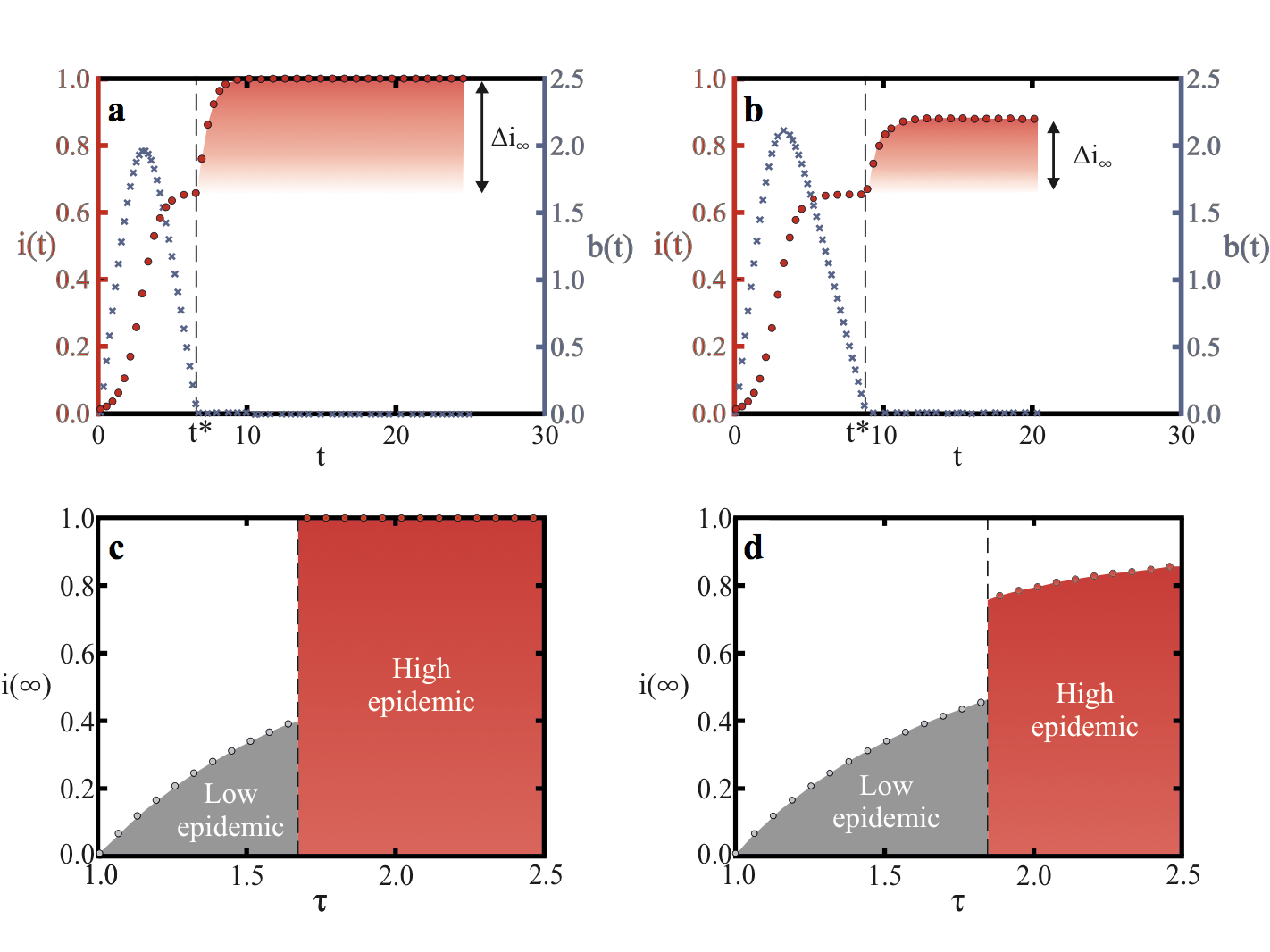}
\caption{\textbf{Generalizing results to the case with a baseline recovery rate independent of budget} \lett{a} Evolution of
the fraction of infected individuals $i(t)$ (red circles) and the 
budget $b(t)$ (blue crosses) in the mean-field model with the same average degree as the friendship network ($\langle k\rangle\approx 8.2355$). All parameters are as in the simulation on the friendship network shown in  Fig.~\ref{fig::school_main}b: most importantly healing costs are above the
critical cost ($c=2>c^*\approx0.833$) and all healing requires a budget contribution ($q_b=0.8$ and $q_0=0$). Clearly the mean-field behavior is very similar to that in the friendship network. \lett{b} Mean-field model with a baseline recovery rate $q_0=0.2$ and $q_b=0.6$ ($q=q_0+q_b=0.8$ as above). The behavior is similar to (a) but $\Delta i_{\infty}$ is smaller due to the baseline recovery. The steady state infection level $i(\infty)$ is bounded from above by $i_{q_0}(\infty)$ (the steady state infection in the pure SIS model with recovery rate $q=q_0=0.2$). \lett{c} The discontinuous phase transition  observed for parameters in (a) as $\tau$ increases (by varying $p$). \lett{d} Similarly, the phase transition observed for the model in panel (b). Although $i(\infty)$ increases continuously with $\tau$ after the phase transition, the jump at $\tau^*$ is  clearly discontinuous.
\label{fig::generalize}}
\end{figure*}

\begin{figure*}
\includegraphics[width=\columnwidth]{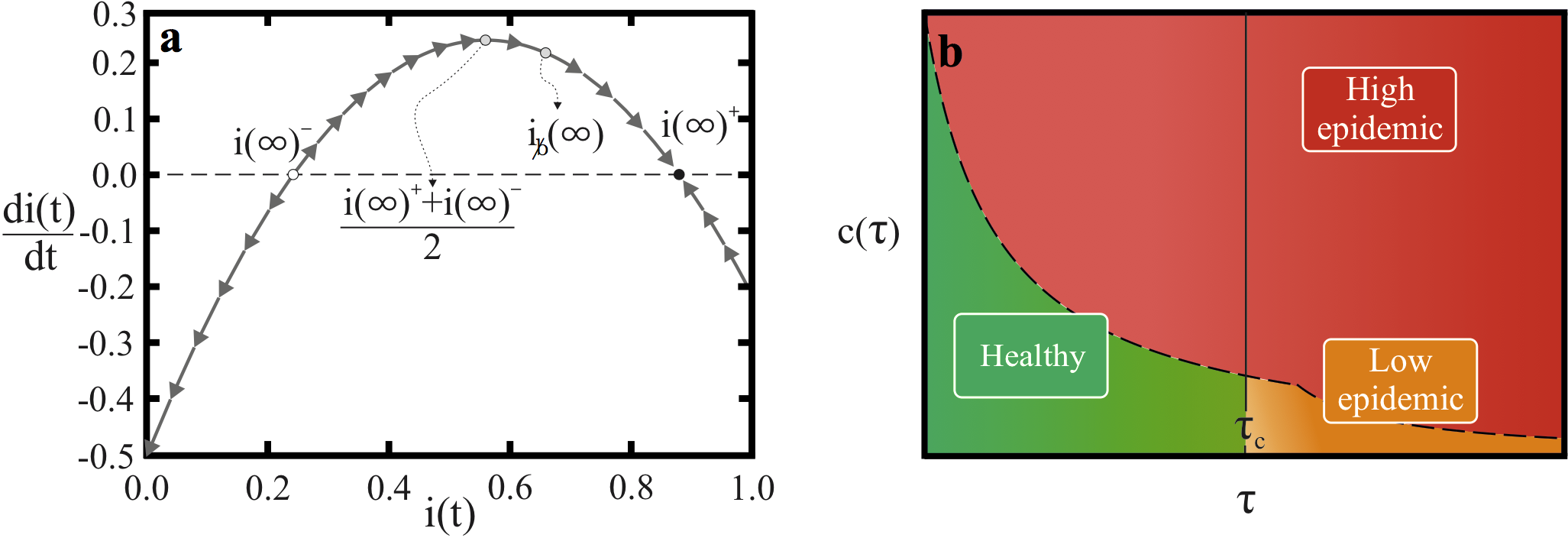}
\caption{\textbf{The explosive epidemic transition to a high epidemic state occurs generically in the bSIS model.} \lett{a} The phase space of Eq.~(\ref{eq::general}) and the corresponding stationary states. In this plot the parameters are the same as in Fig.~\ref{fig::generalize}b,d. The black dots indicate the stationary states of Eq.~(\ref{eq::statstates}) and the arrows illustrate the stability of $i(\infty)^{+}$. In general one can rewrite Eq.~(\ref{eq::general}) to get an equation of a parabola. The equilibrium infection level without the budget restriction is denoted $i_{\not \text{b}}(\infty)$, and it is always the case that $i(\infty)^{+}>i_{\not \text{b}}(\infty)$. \lett{b} The general behavior of the bSIS system can be illustrated by a phase diagram of the different regimes depending on healing costs and basic reproduction number. Three different regimes are found: the
\textit{healthy} regime with a zero fraction of infected individuals
($i(\infty)=0$); the \textit{low epidemic} regime; and the higher infection
\textit{high epidemic} regime. As discussed in the text, transitions between the \textit{healthy} and
\textit{low epidemic} regimes are always continuous, while transitions to the \textit{high epidemic} regime are discontinuous.
\label{fig::stability_fp}}
\end{figure*}

How general the ``explosive epidemic'' that is discontinuous in the model parameters, and that the system behavior depends on the initial conditions, apply over a broad class of models. For instance, the results hold if we allow for a baseline recovery rate $q_0>0$ (see Fig.~\ref{fig::generalize}). Furthermore, we can also show that the results hold for any  budget function $f(b)$ that satisfies the constraints described above. The generality of our results can be easily appreciated if we define $c^*$ as the cost above which the dynamics will converge to a stable equilibrium infection level that is higher than the equilibrium infection level in the model without resource constraints, $i_{\not \text{b}}(\infty)=1-\tau^{-1}$, where $q=q_0+q_b$. We note that in this model infection will always die out, independent of the budget, if $\tau_{q_0}=kp/q_0<1$. Thus, in what follows we only consider the case where $\tau_{q_0}\geq1$.
Solving for the fixed point of the budget by setting $db(t)/dt=0$ in Eq.~(\ref{eq::budget}) and substituting into Eq.~(\ref{eq::infected}) we can express the full infection dynamics using 
\begin{equation}
\label{eq::general}
 \frac{d i(t)}{dt} = k p i(t)[1-i(t)]-c^{-1}[1-i(t)]-q_0i(t).
\end{equation}
As $c$ approaches infinity the resource mediated recovery is no longer possible and this system becomes the SIS model with recovery rate $q_0$, which  has a stable equilibrium that is greater than $1-\tau^{-1}$, namely $1-\tau_{q_0}^{-1}$. However, we can also show that a finite $c^*$ exists. Solving for the fixed points of Eq.~(\ref{eq::general}) we find that there are two fixed points $i(\infty)^+\ge i(\infty)^-$: 
\begin{equation}
\label{eq::statstates}
i(\infty)^{\pm}=\frac{1+c (k p-q_0)\pm \sqrt{[c (q_0-k p)-1]^2-4 c k p}}{2 c k p}.
\end{equation}
We can see from this expression that this equilibrium infection level is bounded above by the equilibrium of the pure SIS model with recovery rate $q_0$ (namely $1-\tau_{q_0}^{-1}$), and approaches it in the limit $c\rightarrow\infty$. However, we can find the critical cost $c^*$ that guarantees $i(\infty)^{+}>i_{\not \text{b}}(\infty)$ by solving for $c^*$ at the fixed point of Eq.~(\ref{eq::general}) and noting that, as in the case $q_0=0$, the smallest possible $c^*$ for a system is given by the stationary point at the largest infection level reached, $i_{max}=\max_t [i_{\not \text{b}}(t)]$. This gives:
\begin{equation}
\label{general_cstar}
c^{\ast} = \frac{1-i_{max}}{i_{max}[k p(1-i_{max})-q]}.
\end{equation}
(In the \textit{Supplementary Information} we present a more detailed characterization of the critical cost $c^*$ that guarantees convergence to the higher infection fixed point.)
Furthermore, it is in fact the case that $i(\infty)^+$ is stable while $i(\infty)^{-}$ is unstable. To characterize the stability we note that Eq.~(\ref{eq::general}) can in general be written as the expression of a parabola with a maximum at $[i(\infty)^{+}+i(\infty)^{-}]/2$, as shown in the phase plot in Fig.~\ref{fig::stability_fp}a.

The size of the jump in infection at this transition, $\Delta i_{\infty}$, is simply the difference between the stationary fraction of infected individuals for the SIS model with resource constraints, $i(\infty)^+$, and that without resource constraints $i_{\not \text{b}}(\infty)$ (see Fig.~\ref{fig::generalize}b). For the special case $q_0=0$ the size of the jump is $\tau^{-1}$ and thus the basic reproduction number fully determines the jump size (see Fig.~\ref{fig::school_main}b and Fig.~\ref{fig::generalize}a). However, for $q_0>0$ there is generally a dependence on the cost $c$.

A key point is that Eq.~(\ref{eq::general}) and all of the stability analysis we have carried out is independent of the budget function $f(b)$. Thus, the argument above also shows that the discontinuous emergence of a new endemic infection level holds under the general conditions for a budget function as previously defined. In the \textit{Supplementary Information} we show a simple example where the discontinuous behavior is observed for a continuous
budget function $f(b)$, further verifying that the discontinuity of the explosive epidemic transition is {\it not} a consequence of the jump in the Heaviside step function at $b=0$. 

\section*{Discussion}

In summary, we have shown that explosive transitions in the epidemic level in a population occur above a
critical healing cost, when the budget for recovery is generated endogenously by the healthy population. Above the critical
healing cost the rate of budget generation cannot cover
the recovery needs. As a consequence, the infection eventually jumps to a higher level than the standard model would predict. The general behavior of the bSIS model can be summarized in a phase
diagram for the mean-field system with two-parameters: healing costs $c$ and
basic reproduction number $\tau$, shown in Figure~\ref{fig::stability_fp}b. Three steady state regimes are found: healthy,
low epidemic, and high epidemic. The healthy regime
converges to a zero fraction of infected individuals. This regime is
observed when $c<c^*$ and $\tau<\tau^*$.
The transition to the
high epidemic regime always leads to a discontinuous jump in the infection incidence, while the transition between
the healthy and epidemic regimes exhibits a continuous change in the infection steady state, as in the classical SIS
model. Importantly, an epidemic can break out even when the standard SIS model predicts that the disease will die out without infecting a significant part of the population. This highlights that to accurately assess the risks posed by a disease outbreak, the constraints that healing resources impose need to be taken into account. Furthermore, in the case that the cost of healing is above the critical cost, a fast response may be necessary to avert an uncontrollable transition into the high epidemic regime.

We emphasize that the behavior we find is very general. We have also shown that it applies across different assumptions about the underlying interaction structure between individuals. Specifically, the scaling exponents describing the transition to the high epidemic regime
in a simulated outbreak, on a real friendship network,
are consistent with our analytical calculations that assume a random interaction network. Furthermore, the discontinuous transition to the
high epidemic regime exists independent of the specific network topology. 
We also show that the details of the budget function do not change the general behavior. Another interesting question is whether our findings hold when we consider other standard epidemiological models. Preliminary work and reasoning indicates that models that introduce a Recovered state, such as the SIR and SIRS models, also display the emergence of an epidemic outbreak when none would be predicted, or the existence of a higher endemic infection level than expected, when the cost of recovery $c$ is high enough. All that would occur is a possible delay as the recovered period grows. These intuitive arguments of course remain to be checked and are an interesting avenue for future work.

Another interesting question is whether our findings extend to budget generation processes where there is a correlation between the ability of an individual to contribute to the budget generation and the likelihood they will become ill. An increased likelihood of falling ill for contributors could more accurately capture the situation of health workers, for example. This refinement is outside of the current scope of our model, but it is easy to see that this correlation would simply make the budget depletion accelerate, rendering the epidemic more explosive. Thus, our model presents the ``best-case'' scenario compared to these more realistic details of budget generation. Similarly, future work could consider the interplay between the heterogeneous
distribution of resource generation and resource demands, across different communities or locations. In this way we could determine how to redistribute the budget, balancing the goal of protecting the greatest number of people in the short-run with the goal of sustaining a high-level of budget contributions. Similarly, we could also determine what locations are strategic points for targeting measures to increase the productivity of healthy individuals.  

Finally, variants of the model we consider may also shed light on non-biological contagion processes where budget constraints have a significant impact on the dynamics of the system. A timely example are bankruptcy cascades
in the banking sector \cite{Battiston07}. Given the high public debt 
levels in most areas of the world, saving banks and
countries suffering from high debt levels might eventually exceed the ability of the 
tax payer to create the necessary resources.

\section*{Acknowledgments}
We would like to thank Alejandro Morales Gallardo for his playful figures of little men, and Michael M\"{a}s, Richard Philip Mann and Lloyd Sanders for their insightful suggestions. We acknowledge financial support from the ETH Risk Center, the
Brazilian institute INCT-SC, ERC Advanced grant number FP7-319968 of the European Research Council, and the Portuguese Foundation for Science and Technology (FCT) under contract no. IF/00255/2013 and PEst-OE/FIS/UI0618/2014. DH and OW acknowledge support by the ERC Advanced Investigator Grant ‘Momentum’ (Grant No. 324247).

\section*{Author contributions statement}

D.H., H.J.H., N.A.M.A and O.W-M. conceived and designed the study, L.B. carried out computational simulations and analytical calculations, O.W-M., L.B. and N.A.M.A wrote the manuscript. All authors reviewed the manuscript. 

\section*{Additional information}

\textbf{Competing financial interests} The author(s) declare no competing financial interests.
\textbf{Data sources} This research uses the public-use dataset from Add Health, a program project designed by J. Richard Udry, Peter S. Bearman, and Kathleen Mullan Harris, and funded by a grant from the National Institute of Child Health and Human Development (P01-HD31921). For data files from Add Health contact Add Health, Carolina Population Center, 123 W. Franklin Street, Chapel Hill, NC 27516-2524, http://www.cpc.unc.edu/addhealth.

\newpage


\begin{thebibliography}{10}
\expandafter\ifx\csname url\endcsname\relax
  \def\url#1{\texttt{#1}}\fi
\expandafter\ifx\csname urlprefix\endcsname\relax\def\urlprefix{URL }\fi
\providecommand{\bibinfo}[2]{#2}
\providecommand{\eprint}[2][]{\url{#2}}

\bibitem{Hufnagel:2004}
\bibinfo{author}{Hufnagel, L.}, \bibinfo{author}{Brockmann, D.} \&
  \bibinfo{author}{Geisel, T.}
\newblock \bibinfo{title}{Forecast and control of epidemics in a globalized
  world}.
\newblock \emph{\bibinfo{journal}{Proc. Natl. Acad. Sci. U.S.A}}
  \textbf{\bibinfo{volume}{101}}, \bibinfo{pages}{15124--15129}
  (\bibinfo{year}{2004}).

\bibitem{Vespignani24072009}
\bibinfo{author}{Vespignani, A.}
\newblock \bibinfo{title}{Predicting the behavior of techno-social systems}.
\newblock \emph{\bibinfo{journal}{Science}} \textbf{\bibinfo{volume}{325}},
  \bibinfo{pages}{425--428} (\bibinfo{year}{2009}).

\bibitem{Colizza06}
\bibinfo{author}{Colizza, V.}, \bibinfo{author}{Barrat, A.},
  \bibinfo{author}{Barthelemy, M.} \& \bibinfo{author}{Vespignani, A.}
\newblock \bibinfo{title}{The role of the airline transportation network in the
  prediction and predictability of global epidemics}.
\newblock \emph{\bibinfo{journal}{Proc. Natl. Acad. Sci. USA}}
  \textbf{\bibinfo{volume}{103}}, \bibinfo{pages}{2015} (\bibinfo{year}{2006}).

\bibitem{Brockmann13}
\bibinfo{author}{Brockmann, D.} \& \bibinfo{author}{Helbing, D.}
\newblock \bibinfo{title}{The hidden geometry of complex, network-driven
  contagion phenomena}.
\newblock \emph{\bibinfo{journal}{Science}} \textbf{\bibinfo{volume}{342}},
  \bibinfo{pages}{1337} (\bibinfo{year}{2013}).

\bibitem{Anderson91}
\bibinfo{author}{Anderson, R.~M.} \& \bibinfo{author}{May, R.~M.}
\newblock \emph{\bibinfo{title}{Infectious diseases of humans: dynamics and
  control}} (\bibinfo{publisher}{Oxford University Press},
  \bibinfo{address}{Oxford New York}, \bibinfo{year}{1991}).

\bibitem{Dorogovtsev08}
\bibinfo{author}{Dorogovtsev, S.~N.}, \bibinfo{author}{Goltsev, A.~V.} \&
  \bibinfo{author}{Mendes, J. F.~F.}
\newblock \bibinfo{title}{Critical phenomena in complex networks}.
\newblock \emph{\bibinfo{journal}{Rev. Mod. Phys.}}
  \textbf{\bibinfo{volume}{80}}, \bibinfo{pages}{1275} (\bibinfo{year}{2008}).

\bibitem{Barrat08}
\bibinfo{author}{Barrat, A.}, \bibinfo{author}{Barth\'elemy, M.} \&
  \bibinfo{author}{Vespignani, A.}
\newblock \emph{\bibinfo{title}{Dynamical processes on complex networks}}
  (\bibinfo{publisher}{Cambridge University Press}, \bibinfo{address}{New
  York}, \bibinfo{year}{2008}).

\bibitem{Kuperman01}
\bibinfo{author}{Kuperman, M.} \& \bibinfo{author}{Abramson, G.}
\newblock \bibinfo{title}{Small world effect in an epidemiological model}.
\newblock \emph{\bibinfo{journal}{Phys. Rev. Lett.}}
  \textbf{\bibinfo{volume}{86}}, \bibinfo{pages}{2909} (\bibinfo{year}{2001}).

\bibitem{Pastor-Satorras01}
\bibinfo{author}{{Pastor-Satorras}, R.} \& \bibinfo{author}{Vespignani, A.}
\newblock \bibinfo{title}{Epidemic spreading in scale-free networks}.
\newblock \emph{\bibinfo{journal}{Phys. Rev. Lett.}}
  \textbf{\bibinfo{volume}{86}}, \bibinfo{pages}{3200} (\bibinfo{year}{2001}).

\bibitem{Newman02}
\bibinfo{author}{Newman, M. E.~J.}
\newblock \bibinfo{title}{Small world effect in an epidemiological model}.
\newblock \emph{\bibinfo{journal}{Phys. Rev. E}} \textbf{\bibinfo{volume}{66}},
  \bibinfo{pages}{016128} (\bibinfo{year}{2002}).

\bibitem{Belik11}
\bibinfo{author}{Belik, V.}, \bibinfo{author}{Geisel, T.} \&
  \bibinfo{author}{Brockmann, D.}
\newblock \bibinfo{title}{Natural human mobility patterns and spatial spread of
  infectious diseases}.
\newblock \emph{\bibinfo{journal}{Physical Review X}}
  \textbf{\bibinfo{volume}{1}}, \bibinfo{pages}{011001} (\bibinfo{year}{2011}).

\bibitem{Dickison12}
\bibinfo{author}{Dickison, M.}, \bibinfo{author}{Havlin, S.} \&
  \bibinfo{author}{Stanley, H.~E.}
\newblock \bibinfo{title}{Epidemics on interconnected networks}.
\newblock \emph{\bibinfo{journal}{Phys. Rev. E}} \textbf{\bibinfo{volume}{85}},
  \bibinfo{pages}{066109} (\bibinfo{year}{2012}).

\bibitem{Kitsak10}
\bibinfo{author}{Kitsak, M.} \emph{et~al.}
\newblock \bibinfo{title}{Identification of influential spreaders in complex
  networks}.
\newblock \emph{\bibinfo{journal}{Nat. Phys.}} \textbf{\bibinfo{volume}{6}},
  \bibinfo{pages}{888} (\bibinfo{year}{2010}).

\bibitem{Balcan10}
\bibinfo{author}{Balcan, D.}, \bibinfo{author}{Gon\c{c}alves, B.},
  \bibinfo{author}{Hu, H.} \& \bibinfo{author}{Ramasco, J.~J.}
\newblock \bibinfo{title}{Modeling the spatial spread of infectious diseases:
  The global epidemic and mobility computational model}.
\newblock \emph{\bibinfo{journal}{Journal of computational Science}}
  \textbf{\bibinfo{volume}{1}}, \bibinfo{pages}{132} (\bibinfo{year}{2010}).

\bibitem{Pastor-Satorras02}
\bibinfo{author}{{Pastor-Satorras}, R.} \& \bibinfo{author}{Vespignani, A.}
\newblock \bibinfo{title}{Immunization of complex networks}.
\newblock \emph{\bibinfo{journal}{Phys. Rev. E}} \textbf{\bibinfo{volume}{65}},
  \bibinfo{pages}{036104} (\bibinfo{year}{2002}).

\bibitem{Cohen01}
\bibinfo{author}{Cohen, R.}, \bibinfo{author}{Havlin, S.} \&
  \bibinfo{author}{{ben-Avraham}, D.}
\newblock \bibinfo{title}{Efficient immunization strategies for computer
  networks and populations}.
\newblock \emph{\bibinfo{journal}{Phys. Rev. Lett.}}
  \textbf{\bibinfo{volume}{91}}, \bibinfo{pages}{247901}
  (\bibinfo{year}{2001}).

\bibitem{Ferguson06}
\bibinfo{author}{Ferguson, N.~M.} \emph{et~al.}
\newblock \bibinfo{title}{Strategies for mitigating an influenza pandemic}.
\newblock \emph{\bibinfo{journal}{Nature}} \textbf{\bibinfo{volume}{442}},
  \bibinfo{pages}{448} (\bibinfo{year}{2006}).

\bibitem{Epstein09}
\bibinfo{author}{Epstein, J.~M.}
\newblock \bibinfo{title}{Modelling to contain pandemics}.
\newblock \emph{\bibinfo{journal}{Nature}} \textbf{\bibinfo{volume}{460}},
  \bibinfo{pages}{687} (\bibinfo{year}{2009}).

\bibitem{Schneider11}
\bibinfo{author}{Schneider, C.~M.}, \bibinfo{author}{Mihaljev, T.},
  \bibinfo{author}{Havlin, S.} \& \bibinfo{author}{Herrmann, H.~J.}
\newblock \bibinfo{title}{Suppressing epidemics with a limited amount of
  immunization units}.
\newblock \emph{\bibinfo{journal}{Phys. Rev. E}} \textbf{\bibinfo{volume}{84}},
  \bibinfo{pages}{061911} (\bibinfo{year}{2011}).

\bibitem{vandenBroeck11}
\bibinfo{author}{\mbox{van den Broeck}, W.} \emph{et~al.}
\newblock \bibinfo{title}{The gleamviz computational tool, a publicly available
  software to explore realistic epidemic spreading scenarios at the global
  scale}.
\newblock \emph{\bibinfo{journal}{BMC Infectious Diseases}}
  \textbf{\bibinfo{volume}{11}}, \bibinfo{pages}{37} (\bibinfo{year}{2011}).

\bibitem{Colizza07}
\bibinfo{author}{Colizza, V.}, \bibinfo{author}{Barrat, A.},
  \bibinfo{author}{Barthelemy, M.}, \bibinfo{author}{Valleron, A.-J.} \&
  \bibinfo{author}{Vespignani, A.}
\newblock \bibinfo{title}{Modeling the worldwide spread of pandemic influenza:
  baseline case and containment interventions}.
\newblock \emph{\bibinfo{journal}{\mbox{PLoS} Medicine}}
  \textbf{\bibinfo{volume}{4}}, \bibinfo{pages}{e13} (\bibinfo{year}{2007}).

\bibitem{Helbing13}
\bibinfo{author}{Helbing, D.}
\newblock \bibinfo{title}{Globally networked risks and how to respond}.
\newblock \emph{\bibinfo{journal}{Nature}} \textbf{\bibinfo{volume}{497}},
  \bibinfo{pages}{51} (\bibinfo{year}{2013}).

\bibitem{gallup2001economic}
\bibinfo{author}{Gallup, J.~L.} \& \bibinfo{author}{Sachs, J.~D.}
\newblock \bibinfo{title}{The economic burden of malaria}.
\newblock \emph{\bibinfo{journal}{The American journal of tropical medicine and
  hygiene}} \textbf{\bibinfo{volume}{64}}, \bibinfo{pages}{85--96}
  (\bibinfo{year}{2001}).

\bibitem{sachs2002economic}
\bibinfo{author}{Sachs, J.} \& \bibinfo{author}{Malaney, P.}
\newblock \bibinfo{title}{The economic and social burden of malaria}.
\newblock \emph{\bibinfo{journal}{Nature}} \textbf{\bibinfo{volume}{415}},
  \bibinfo{pages}{680--685} (\bibinfo{year}{2002}).

\bibitem{bell2003long}
\bibinfo{author}{Bell, C.}, \bibinfo{author}{Devarajan, S.} \&
  \bibinfo{author}{Gersbach, H.}
\newblock \bibinfo{title}{The long-run economic costs of aids: Theory and an
  application to {S}outh {A}frica}.
\newblock \emph{\bibinfo{journal}{World Bank Policy Research Working Paper}}
  (\bibinfo{year}{2003}).

\bibitem{mclean2005sars}
\bibinfo{author}{McLean, A.~R.}, \bibinfo{author}{May, R.~M.},
  \bibinfo{author}{Pattison, J.}, \bibinfo{author}{Weiss, R.} \emph{et~al.}
\newblock \emph{\bibinfo{title}{SARS: a case study in emerging infections.}}
  (\bibinfo{publisher}{Oxford University Press}, \bibinfo{year}{2005}).

\bibitem{webby2003we}
\bibinfo{author}{Webby, R.~J.} \& \bibinfo{author}{Webster, R.~G.}
\newblock \bibinfo{title}{Are we ready for pandemic influenza?}
\newblock \emph{\bibinfo{journal}{Science}} \textbf{\bibinfo{volume}{302}},
  \bibinfo{pages}{1519--1522} (\bibinfo{year}{2003}).

\bibitem{pandey2014strategies}
\bibinfo{author}{Pandey, A.} \emph{et~al.}
\newblock \bibinfo{title}{Strategies for containing ebola in west africa}.
\newblock \emph{\bibinfo{journal}{Science}} \textbf{\bibinfo{volume}{346}},
  \bibinfo{pages}{991--995} (\bibinfo{year}{2014}).

\bibitem{merler2015spatiotemporal}
\bibinfo{author}{Merler, S.} \emph{et~al.}
\newblock \bibinfo{title}{Spatiotemporal spread of the 2014 outbreak of ebola
  virus disease in liberia and the effectiveness of non-pharmaceutical
  interventions: a computational modelling analysis}.
\newblock \emph{\bibinfo{journal}{The Lancet Infectious Diseases}}
  \textbf{\bibinfo{volume}{15}}, \bibinfo{pages}{204--211}
  (\bibinfo{year}{2015}).

\bibitem{Parshani10}
\bibinfo{author}{Parshani, R.}, \bibinfo{author}{Carmi, S.} \&
  \bibinfo{author}{Havlin, S.}
\newblock \bibinfo{title}{Epidemic threshold for the
  susceptible-infectious-susceptible model on random networks}.
\newblock \emph{\bibinfo{journal}{Phys. Rev. Lett.}}
  \textbf{\bibinfo{volume}{104}}, \bibinfo{pages}{258701}
  (\bibinfo{year}{2010}).

\bibitem{Boguna13}
\bibinfo{author}{Boguna, M.}, \bibinfo{author}{Castellano, C.} \&
  \bibinfo{author}{Pastor-Satorras, R.}
\newblock \bibinfo{title}{Nature of the epidemic threshold for the
  susceptible-infected-susceptible dynamics in networks}.
\newblock \emph{\bibinfo{journal}{Phys. Rev. Lett.}}
  \textbf{\bibinfo{volume}{111}}, \bibinfo{pages}{068701}
  (\bibinfo{year}{2013}).

\bibitem{Marro99}
\bibinfo{author}{Marro, J.} \& \bibinfo{author}{Dickman, R.}
\newblock \emph{\bibinfo{title}{Nonequilibrium phase transitions in lattice
  models}} (\bibinfo{publisher}{Cambridge University Press},
  \bibinfo{address}{New York}, \bibinfo{year}{1999}).

\bibitem{Watts98}
\bibinfo{author}{Watts, D.~J.} \& \bibinfo{author}{Strogatz, S.~H.}
\newblock \bibinfo{title}{Collective dynamics of small-world networks}.
\newblock \emph{\bibinfo{journal}{Nature}} \textbf{\bibinfo{volume}{393}},
  \bibinfo{pages}{440} (\bibinfo{year}{1998}).

\bibitem{Palla05}
\bibinfo{author}{Palla, G.}, \bibinfo{author}{Der{\'e}nyi, I.},
  \bibinfo{author}{Farkas, I.} \& \bibinfo{author}{Vicsek, T.}
\newblock \bibinfo{title}{Uncovering the overlapping community structure of
  complex networks in nature and society}.
\newblock \emph{\bibinfo{journal}{Nature}} \textbf{\bibinfo{volume}{435}},
  \bibinfo{pages}{814} (\bibinfo{year}{2005}).

\bibitem{Easley10}
\bibinfo{author}{Easley, D.} \& \bibinfo{author}{Kleinberg, J.}
\newblock \emph{\bibinfo{title}{Networks, crowds, and markets}}
  (\bibinfo{publisher}{Cambridge University Press}, \bibinfo{address}{New
  York}, \bibinfo{year}{2010}).

\bibitem{newman2011communities}
\bibinfo{author}{Newman, M.}
\newblock \bibinfo{title}{Communities, modules and large-scale structure in
  networks}.
\newblock \emph{\bibinfo{journal}{Nature Physics}}
  \textbf{\bibinfo{volume}{8}}, \bibinfo{pages}{25--31} (\bibinfo{year}{2011}).

\bibitem{Fortunato10}
\bibinfo{author}{Fortunato, S.}
\newblock \bibinfo{title}{Community detection in graphs}.
\newblock \emph{\bibinfo{journal}{Phys. Rep.}} \textbf{\bibinfo{volume}{486}},
  \bibinfo{pages}{75} (\bibinfo{year}{2010}).

\bibitem{Moreira96}
\bibinfo{author}{Moreira, A.~G.} \& \bibinfo{author}{Dickman, R.}
\newblock \bibinfo{title}{Critical dynamics of the contact process with
  quenched disorder}.
\newblock \emph{\bibinfo{journal}{Phys. Rev. E}} \textbf{\bibinfo{volume}{54}},
  \bibinfo{pages}{R3090} (\bibinfo{year}{1996}).

\bibitem{Battiston07}
\bibinfo{author}{Battiston, S.}, \bibinfo{author}{Gatti, D.~D.},
  \bibinfo{author}{Gallegati, M.}, \bibinfo{author}{Greenwald, B.} \&
  \bibinfo{author}{Stiglitz, J.~E.}
\newblock \bibinfo{title}{Credit chains and bankruptcy propagation in
  production networks}.
\newblock \emph{\bibinfo{journal}{J. Econ. Dyn. Control}}
  \textbf{\bibinfo{volume}{31}}, \bibinfo{pages}{2061--2084}
  (\bibinfo{year}{2007}).

\bibitem{Gonzalez07}
\bibinfo{author}{Gonz{\'a}lez, M.~C.}, \bibinfo{author}{Herrmann, H.~J.},
  \bibinfo{author}{Kert{\'e}sz, J.} \& \bibinfo{author}{Vicsek, T.}
\newblock \bibinfo{title}{Community structure and ethnic preference in school
  friendship networks}.
\newblock \emph{\bibinfo{journal}{Physica A}} \textbf{\bibinfo{volume}{379}},
  \bibinfo{pages}{307} (\bibinfo{year}{2007}).

\end{thebibliography}
\end{document}